\documentclass[a4paper]{article}

\usepackage{amsmath}
\usepackage{amsfonts}
\usepackage{latexsym}
\usepackage{feynmp}

\newlength{\authorspace}
\newlength{\preprintspace}
\begin{document}

\title{\bf Hamilton-Jacobi Equation and the \\
Tree Formula for Proper Vertices}
\author{\Large Chr. Wieczerkowski \\[\authorspace]
Inst.~f.~Theor.~Physik I, Universit\"at M\"unster, \\ 
Wilhelm-Klemm-Str.~9, 48159 M\"unster, Germany}
\date{September 24, 1998}
\maketitle
\vspace{-\preprintspace}
   \hfill MS-TPI-98-23 
\vspace{\preprintspace}

\section{Introduction}

Proper vertices are central elements of a Statistical Field 
Theory. See \cite{Zinn-Justin:1990} and references therein.
They are related to the connected vertices (or correlators) 
through a tree formula. This fact is both well known and 
ubiquitously used in the literature. In pedagogical accounts, 
the tree formula is usually motivated by a low order 
computation (in powers of fields) but neither written down 
explicitely nor proved to all orders. In this note, we intend 
to fill this pedagogical gap. Our proof is an elaboration of 
a remark in \cite{Coleman:1973}. The proof is complete aside 
of convexity and stability issues needed to have well defined 
Legendre transforms and path integrals. Trivial normalization 
constants will be neglected.

The generating functions of the proper vertices and the connected 
vertices are Legendre transforms of each other:
\begin{equation}
\Gamma(\Phi)=
\sup_{J} 
\biggl\{
   (\Phi,J)-W(J)
\biggr\}
\label{1}
\end{equation}
For definiteness, we consider the case of a real scalar field 
theory on a finite lattice $\Lambda\subset\mathbb{Z}^{D}$, where 
$\Phi,J\in\{\phi:\Lambda\rightarrow\mathbb{R}\}$ and 
$(\Phi,J)=\sum_{x\in\Lambda}\Phi(x)J(x)$. In Statistical Field
Theory, the generating function  $W(J)$ comes as a path integral 
\begin{equation}
W(J)=
\hbar\log
\left[
   \int\mathcal{D}\phi\;
   \exp\left[
      \frac{1}{\hbar}
      \biggl\{
         (\phi,J)
         -S(\phi)
      \biggr\}
   \right]
\right]
\label{2}
\end{equation}
in terms of the classical action $S(\phi)$.
In the classical limit, when $\hbar\rightarrow 0$, the path integral
becomes a Legendre transform. And, since the Legendre transform
is involutive, (\ref{1}) inverts the classical limit of (\ref{2}). 
The classical limit of $\Gamma(\Phi)$ is thus $S(\phi)$.
The tree formula follows from an analogous path integral 
representation for the inverse of (\ref{1}),
\begin{equation}
W(J)=
\lim_{\hbar\rightarrow 0}\hbar\log
\left[
   \int\mathcal{D}\Phi\;
   \exp\left[
      \frac{1}{\hbar}
      \biggl\{
         (\Phi,J)
         -\Gamma(\Phi)
      \biggr\}
   \right]
\right].
\label{3}
\end{equation}
Of course, $\hbar$ is here only an auxiliary variable cousin of 
the constant of nature. One then writes $\Gamma(\Phi)=
\frac{1}{2}(\Phi,G^{-1}\Phi)+\Gamma_{1}(\Phi)$ and 
performs a perturbation expansion. The propagator is $G$, the 
interaction is $\Gamma_{1}(\Phi)$. Only tree diagrams 
contribute. To lowest orders (in powers of $\Gamma_{1}(\Phi)$), 
this perturbation expansion reads
\begin{gather}
W(J)=
\frac{1}{2}(J,GJ)
-\Gamma_{1}(GJ)
+\frac{1}{2}\sum_{x_{1},y_{1}\in\Lambda}\;G(x_{1},y_{1})\;
\Gamma_{1,x_{1}}(G J)\;\Gamma_{1,y_{1}}(GJ)
\nonumber\\
-\frac{1}{3!}\sum_{x_{1},y_{1},x_{2},y_{2}\in\Lambda}
G(x_{1},y_{1})\;G(x_{2},y_{2})\;
\biggl\{
   \Gamma_{1,x_{1}}(G J)\;
   \Gamma_{1,y_{1}x_{2}}(G J)\;
   \Gamma_{1,y_{2}}(G J)
\nonumber\\
+
   \Gamma_{1,x_{1}x_{2}}(G J)\;
   \Gamma_{1,y_{1}}(G J)\;
   \Gamma_{1,y_{2}}(G J)
+
   \Gamma_{1,x_{1}}(G J)\;
   \Gamma_{1,x_{2}}(G J)\;
   \Gamma_{1,y_{1}y_{2}}(G J)
\biggr\}
\nonumber\\
+O(\Gamma_{1}^{4})
\label{4}
\end{gather}
where 
\begin{equation}
\Gamma_{1,x_{1}\cdots x_{n}}(\Phi)=
\frac{\partial^{n}}{\partial\Phi(x_{1})\cdots\partial\Phi(x_{n})}
\Gamma_{1}(\Phi).
\label{5}
\end{equation}
We will show that this series, beginning as in (\ref{4}), is the 
iterative solution to a Hamilton-Jacobi equation. 
We will then write down explicitely the terms of any order. 
With this expression in our hands, we will demonstrate that the 
tree expansion converges. 

\section{Hamilton-Jacobi Equation}

Our first step is to write (\ref{3}) in a fancy way. From the 
fancy formula, we deduce a Hamilton-Jacobi differential equation. 
More details (and references) thereabout can be found in
\cite{Brydges:1992}.

\subsection{Propagator $\boldsymbol{G}$}

The tree formula involves lines. These lines are propagators $G$. 
They correspond to quadratic (or kinetic) terms in $\Gamma(\Phi)$ 
and $W(J)$, which are brotherly split from $\Gamma(\Phi)=
\Gamma_{0}(\Phi)+\Gamma_{1}(\Phi)$ and $W(J)=W_{0}(J)+W_{1}(J)$: 
\begin{equation}
\Gamma_{0}(\Phi)=
\frac{1}{2} (\Phi,G^{-1}\Phi), 
\qquad
W_{0}(J)=
\frac{1}{2} (J,G J).
\label{6}
\end{equation}
Usually, but not necessarily, $W_{0}(J)$ is the full quadratic term 
of $W(J)$, and $G(x,y)$ is the (full) connected two point function 
of the theory:
\begin{equation}
G(x,y)=
\frac{\partial^{2}}{\partial J(x)\partial J(y)}
W(J)\bigg\vert_{J=0}
\label{7}
\end{equation}
However, there is nothing wrong with leaving behind a quadratic 
remainder in $W_{1}(J)$, which then adds a quadratic interaction.
To be on safe ground, we will assume that $G(x,y)$ is real, 
symmetric, and (that $W_{0}(J)$ is) positive. Then $G$ is 
invertible, and this inverse appears in $\Gamma_{0}(\Phi)$. By 
completing a square, (\ref{3}) becomes
\begin{equation}
W_{1}(J)=
\lim_{\hbar\rightarrow 0}
\hbar\log
\left[
   \int\mathrm{d}\mu_{\hbar G}(\Phi)\;
   \exp\left\{
      -\frac{1}{\hbar} \Gamma_{1}(\Phi+G J)
   \right\}
\right]
\label{8}
\end{equation}
where 
\begin{equation}
\mathrm{d}\mu_{G}(\Phi)=
\det(2\pi G)^{-\frac{1}{2}}\;
\mathcal{D}\Phi\;
\exp\left\{
   -\frac{1}{2}(\Phi,G^{-1}\Phi)
\right\}
\label{9}
\end{equation}
is the Gaussian measure on field space associated with the 
propagator $G$. In (\ref{8}), the external source $J$ is dressed
with the propagator $G$. It is convenient to introduce a 
new symbol for this dressed source: $\Psi=G J$. Formula (\ref{8})
is the optimal starting point for the perturbation expansion
(\ref{4}).

\subsection{Interpolation parameter}

Our second step is to tune up $\hbar$ temporarily to finite
values in (\ref{8}). Moreover, we join to $\hbar$ a second 
interpolation parameter $t$, which takes values in between zero 
and one:
\begin{equation}
W_{1}(\Psi,t)=
\hbar\log
\left[
   \int\mathrm{d}\mu_{t\hbar G}(\Phi)\;
   \exp\left\{
      -\frac{1}{\hbar} \Gamma_{1}(\Phi+\Psi)
   \right\}
\right]
\label{10}
\end{equation}
A look at (\ref{8}) reveals that $\hbar$ appears at two places
in the integral. The idea with the new interpolation parameter is
to tune both $\hbar$s independently. The interpolated quantity
solves the differential equation
\begin{equation}
\frac{\partial}{\partial t}W_{1}(\Psi,t)=
\frac{1}{2}\;
\left(
   \frac{\partial}{\partial\Psi} W_{1}(\Psi,t), G\,
   \frac{\partial}{\partial\Psi} W_{1}(\Psi,t)
\right)
-\frac{\hbar}{2}\;
\left(
   \frac{\partial}{\partial\Psi}, G\,
   \frac{\partial}{\partial\Psi}
\right)\;
W_{1}(\Psi,t)
\label{11}
\end{equation}
with the initial condition $W_{1}(\Psi,0)=-\Gamma_{1}(\Psi)$.
The aim of the game is $W_{1}(\Psi,1)$ at $\hbar=0$. 
But (\ref{11}) is a perfect instant to put $\hbar=0$, 
remaining with only one interpolation parameter $t$. The 
result is the Hamilton-Jacobi 
\begin{equation}
\frac{\partial}{\partial t}W_{1}(\Psi,t)=
\frac{1}{2}\;
\left(
   \frac{\partial}{\partial\Psi} W_{1}(\Psi,t), G\,
   \frac{\partial}{\partial\Psi} W_{1}(\Psi,t)
\right).
\label{12}
\end{equation}
The initial condition is independent of $\hbar$. Therefore 
(\ref{12}) is integrated to
\begin{equation}
W_{1}(\Psi,t)=
\frac{1}{2}\;
\int_{0}^{t}\mathrm{d}s\;
\left(
   \frac{\partial}{\partial\Psi} W_{1}(\Psi,s), G\,
   \frac{\partial}{\partial\Psi} W_{1}(\Psi,s)
\right)
-\Gamma_{1}(\Psi)
\label{13}
\end{equation}
Remarkably, the solution (\ref{13}) to the Hamilton-Jacobi
equation (\ref{12}) performs a Legendre transformation of 
the initial data. The Legendre transform is recovered as the 
boundary value
\begin{equation}
W_{1}(J)=
W_{1}(GJ,1)
\label{14}
\end{equation}
The iteration of (\ref{13}) generates a tree formula. Indeed, 
the bilinear term creates a branch in every iteration step. 

\section{Tree expansion}

A basic version of the tree expansion is obtained as follows.
We develop $W_{1}(\Psi,t)$ into a power series in $t$,
\begin{equation}
W_{1}(\Psi,t)=
\sum_{n=0}^{\infty}
\frac{t^{n}}{n!}\;W_{1}^{(n)}(\Psi).
\label{15}
\end{equation}
The order zero is our initial condition 
$W_{1}^{(0)}(\Psi)=-\Gamma_{1}(\Psi)$. 
The higher orders are recursively determined by 
\begin{equation}
W_{1}^{(n+1)}(\Psi)=
\frac{1}{2}\sum_{m=0}^{n}
\binom{n}{m}\;
\left(
   \frac{\partial}{\partial\Psi}W_{1}^{(m)}(\Psi),G\,
   \frac{\partial}{\partial\Psi}W_{1}^{(n-m)}(\Psi)
\right).
\label{16}
\end{equation}
The illustrative formula (\ref{4}) is found by iterating 
(\ref{16}) a few times. The outcome of an infinite iteration of 
(\ref{16}) is the renowned tree formula
\begin{equation}
W_{1}^{(n)}(\Psi)=
\frac{(-1)^{n+1}}{n+1}\sum_{\tau\in T_{n}(\{1,2,\ldots,n+1\})}
\bigtriangleup_{\Psi}(\tau)\;
\prod_{m=1}^{n+1}
\Gamma(\Psi_{m})\bigg\vert_{\Psi_{1}=\cdots=\Psi_{n+1}=\Psi}
\label{17}
\end{equation}
Here $T_{n}(\{1,2,\ldots,n+1\})$ denotes the set of 
$n$-trees on the set $\{1,2,\ldots,n+1\}$. An $n$-tree is 
a set of links
\begin{equation}
\tau=
\bigl\{
   \{i_{1},j_{1}\},\cdots,\{i_{n},j_{n}\}
\bigr\},
\label{18}
\end{equation}
consisting of $n$ sets $\{i,j\}$ (the links) with 
$i,j\in\{1,2,\ldots,n+1\}$ and $i\neq j$ such that 
\begin{equation}
\{1,2,\ldots,n+1\}=
\bigcup_{a=1}^{n}\{i_{a},j_{a}\}
\label{19}
\end{equation}
A less formal explanation is this: draw $n+1$ dots on a 
sheet of paper; connect the dots pairwise with $n$ links 
such that the resulting diagram is connected; then this is
a tree diagram. If you happen to draw a loop, the resulting
diagram will be disconnected.
With each $n$-tree is associated a linking operator
\begin{equation}
\bigtriangleup_{\Psi}(\tau)=
\prod_{a=1}^{n}
\left(
   \frac{\partial}{\partial_{\Psi_{i_{a}}}},G\;
   \frac{\partial}{\partial_{\Psi_{j_{a}}}}
\right).
\label{20}
\end{equation}
As its name reveals, the linking operator links together 
the product of proper vertices in (\ref{17}).

\subsection{Examples}

There are three $2$-trees on $\{1,2,3\}$:
\begin{gather*}
\bigl\{\{1,2\},\{2,3\}\bigr\},\quad
\bigl\{\{1,3\},\{2,3\}\bigr\},\quad
\bigl\{\{1,2\},\{1,3\}\bigr\}.
\end{gather*}
There are sixteen $3$-trees on $\{1,2,3,4\}$:
twelve straight trees
\begin{gather*}
\bigl\{\{1,2\},\{2,3\},\{3,4\}\bigr\},\quad
\bigl\{\{1,2\},\{2,3\},\{1,4\}\bigr\},\\
\bigl\{\{1,2\},\{1,3\},\{2,4\}\bigr\},\quad
\bigl\{\{1,2\},\{1,4\},\{3,4\}\bigr\},\\
\bigl\{\{1,3\},\{1,4\},\{2,3\}\bigr\},\quad
\bigl\{\{1,3\},\{2,3\},\{2,4\}\bigr\},\\
\bigl\{\{1,2\},\{1,3\},\{3,4\}\bigr\},\quad
\bigl\{\{1,4\},\{2,3\},\{3,4\}\bigr\},\\
\bigl\{\{1,3\},\{2,4\},\{3,4\}\bigr\},\quad
\bigl\{\{1,2\},\{1,4\},\{3,4\}\bigr\},\\
\bigl\{\{1,4\},\{2,3\},\{2,4\}\bigr\},\quad
\bigl\{\{1,2\},\{1,4\},\{2,4\}\bigr\},
\end{gather*}
and four trees with a fork
\begin{gather*}
\bigl\{\{1,2\},\{1,3\},\{1,4\}\bigr\},\quad
\bigl\{\{2,1\},\{2,3\},\{2,4\}\bigr\},\\
\bigl\{\{3,1\},\{3,2\},\{3,4\}\bigr\},\quad
\bigl\{\{4,1\},\{4,2\},\{4,3\}\bigr\}.
\end{gather*}

\subsection{Proof by Induction on $\boldsymbol{n}$}

The trivial tree is only a single dot. It suffices therefore
to prove the induction step $n\rightarrow n+1$. From the 
induction hypothesis, we find 
\begin{gather}
W_{1}^{(n+1)}(\Psi)=
\frac{(-1)^{n+2}}{2\,(n+1)\,(n+2)}\sum_{m=0}^{n}
\binom{n+2}{m+1}
\nonumber\\
\sum_{i=1}^{m+1}\sum_{j=m+2}^{n+2}
\sum_{\tau_{1}\in T_{m}\bigl(\{1,2,\ldots,m+1\}\bigr)}
\sum_{\tau_{2}\in T_{n-m}\bigl(\{m+2,m+3,\ldots,n+2\}\bigr)}
\nonumber\\
\left(
   \frac{\partial}{\partial\Psi_{i}},G\,
   \frac{\partial}{\partial\Psi_{j}}
\right)\;
\bigtriangleup_{\Psi}(\tau_{1})\;
\bigtriangleup_{\Psi}(\tau_{2})\;
\prod_{i=1}^{n+2}\Gamma(\Psi_{i})
\bigg\vert_{\Psi_{1}=\cdots=\Psi_{n+2}=\Psi}
\label{21}
\end{gather}
But $\binom{n+2}{m+1}$ is just the number of subsets of 
$\{1,2,\ldots,n+2\}$ with $m+1$ elements. Therefore, 
\begin{gather}
W_{1}^{(n+1)}(\Psi)=
\frac{(-1)^{n+2}}{2\,(n+1)\,(n+2)}\sum_{m=0}^{n}
\sum_{\{1,2,\ldots,n+2\}=I\cup J}
\delta_{\vert I\vert,m+1}\,\delta_{\vert J\vert, n-m+1}
\nonumber\\
\sum_{i\in I}\sum_{j\in J}
\sum_{\tau_{1}\in T_{m}(I)}
\sum_{\tau_{2}\in T_{n-m}(J)}
\left(
   \frac{\partial}{\partial\Psi_{i}},G\,
   \frac{\partial}{\partial\Psi_{j}}
\right)\;
\bigtriangleup_{\Psi}(\tau_{1})\;
\bigtriangleup_{\Psi}(\tau_{2})\;
\nonumber\\
\prod_{i=1}^{n+2}\Gamma(\Psi_{i})
\bigg\vert_{\Psi_{1}=\cdots=\Psi_{n+2}=\Psi}
\label{22}
\end{gather}
The factor $\frac{1}{2}$ takes care of the fact that $(I,J)$ 
and $(J,I)$ yield equal contributions. The factor $\frac{1}{n+1}$
takes care of the fact that any $(n+1)$-tree on $\{1,2,\ldots,n+2\}$
has (tautologically) $n+1$ links, and can therefore be broken 
up into $n+1$ pairs of subtrees by cutting a line. Eq.~(\ref{22})
conversely assembles $(n+1)$-trees in all of these possible 
ways. The result is
\begin{gather}
W_{1}^{(n+1)}(\Psi)=
\frac{(-1)^{n+2}}{n+2}
\sum_{\tau\in T_{n+1}\bigl(\{1,2,\ldots,n+2\}\bigr)}
\bigtriangleup_{\Psi}(\tau)\;
\prod_{i=1}^{n+2}\Gamma(\Psi_{i})
\bigg\vert_{\Psi_{1}=\cdots=\Psi_{n+2}=\Psi}
\label{23}
\end{gather}
The induction step is complete. $\Box$

The result of the tree expansion is the following tree 
formula for the Legendre transform:
\begin{equation}
W(J)=
\frac{1}{2}(J,G J)
+\sum_{n=1}^{\infty}\frac{(-1)^{n}}{n!}
\sum_{\tau\in T_{n-1}\bigl(1,2,\ldots,n\bigr)}
\bigtriangleup_{\Psi}(\tau) 
\prod_{i=1}^{n}\Gamma_{1}(\Psi_{i})
\bigg\vert_{\Psi_{1}=\cdots=\Psi_{n}=G J}
\label{24}
\end{equation}
Quite remarkably, the different looking formulas (\ref{1}), 
(\ref{13}), and (\ref{24}) are variants of the same Legendre 
transform. 

\section{Convergence of the Tree Expansion}

There remains the question whether (\ref{24}) is formal 
expression. In perturbation theory, this question is 
irrelevant since always only finitely many terms in the 
tree expansion contribute to finite order expressions in
perturbation theory. But the perturbation (or loop) 
expansion is another expansion on top of the tree expansion.  

The Hamilton-Jacobi equation (\ref{12}) is an ideal tool 
to study the convergence of the tree expansion. The first
thing to do is to choose a suitable norm. A very suitable 
norm is 
\begin{gather}
\| W_{1}(\cdot,t)\|_{h}=
\sum_{n=0}^{\infty}\frac{h^{n}}{n!}\;
\sup_{x_{1}\in\Lambda}\sum_{x_{2},\ldots,x_{n}\in\Lambda}\;
\vert W_{1,x_{1}\ldots x_{n}}(0,t)\vert.
\label{25}
\end{gather}
The meaning of $h$ is a radius of analyticity in source
space. The point with this norm is that (\ref{12}) 
immediately implies a differential inequality, namely
\begin{equation}
\frac{\partial}{\partial t}\|W_{1}(\cdot,t)\|_{h}
\leq
\frac{\| G\|}{2}\;
\left(
   \frac{\partial}{\partial h}\|W_{1}(\cdot,t)\|_{h}
\right)^{2},
\label{26}
\end{equation}
where
\begin{equation}
\|G\|=\sup_{x\in\Lambda}\sum_{y\in\Lambda}
\vert G(x,y)\vert.
\label{27}
\end{equation}
We will assume that $\| G\|$ is finite. (On a finite lattice, 
this condition is per se fulfilled.) Any solution
of (\ref{26}) is majorized by the solution to the 
simpler Hamilton-Jacobi equation
\begin{equation}
\frac{\partial}{\partial t}f(t,h)=
\frac{\| G\|}{2}\;
\left(
   \frac{\partial}{\partial h}f(t,h)
\right)^{2}.
\label{28}
\end{equation}
The initial condition $W_{1}(\Psi,0)=-\Gamma_{1}(\Psi)$ 
gives us the initial condition $f(0,h)=\|\Gamma_{1}\|_{h}$
for (\ref{28}). Depending on the properties of this 
initial value, one ascertains the existence (and an estimate
on) the boundary value $f(1,h)$. Let us only quote the 
following result from \cite{Brydges/Yau:1990}:

If $\|\Gamma_{1}\|_{h}$ is holomorphic on a disc $\{h\in\mathbb{C}:
\vert h\vert\leq h_{0}\}$, and if $\|\Gamma_{1}\|_{h_{0}}\leq
\frac{(h_{0}-h_{1})^{2}}{16\,\|G\|}$ for some $h_{1}\in (0,h_{0})$, 
then $\|W_{1}(\cdot,1)\|_{h}$ is holomorphic on the (smaller)
disc $\{h\in\mathbb{C}:\vert h\vert\leq h_{1}\}$, and it 
satisfies the bound $\|W_{1}(\cdot,1)\|_{h_{1}}\leq
\|\Gamma_{1}\|_{h_{0}}$.
 
\section{Concluding Remarks}

We have shown that the Legendre transform can be advantageously 
studied as the solution of a Hamilton-Jacobi equation. 
Both the tree formula and an estimate on the tree expansion 
are straight forward consequences of the Hamilton-Jacobi 
equation. 

Needless to say that this method also applies to 
field theories with a more complicated field content. 
It is immediately applicable to the case of Fermions. 
With some extra effort, all formulas can be extended to the 
continuum limit. 

Since the Legendre transform is involutive, one immediately 
also has a tree formula for $\Gamma(\Phi)$ in terms of 
$W(J)$. All that has to be changed is that $G$ has to be 
replaced by its inverse $G^{-1}$. 

Although there are good reasons to study the proper vertices 
in Statistical Field Theory, modifications of the 
Hamilton-Jacobi formulation of the Legendre transform might
prove to be useful in other contexts.

\section*{Graphical representation}

The Hamilton-Jacobi equation (\ref{12}) has the following 
graphical representation (expanded in powers of fields):

\begin{fmffile}{graph}
\fmfset{dot_size}{2pt}
\begin{gather*}
\frac{\partial}{\partial t}
\parbox{60pt}{
\begin{fmfgraph*}(90,55)
\fmfincoming{i1,i2,i3,i4,i5}
\fmfoutgoing{o1,o2,o3,o4,o5}
\fmfv{d.sh=circle,d.size=30pt,d.filled=empty}{v1}
\fmf{phantom}{i2,h1,hh1,v1,hh2,h2,i3}
\fmf{phantom}{i4,h3,hh3,v1,hh4,h4,o2}
\fmf{phantom}{o3,h5,hh5,v1,hh6,h6,o4}
\fmfdotn{h}{3}
\fmf{dashes,tension=1.4}{i1,v1,i5}
\fmf{phantom}{o1,v1,o5}
\fmflabel{$1$}{i5}
\fmflabel{$n$}{i1}
\end{fmfgraph*}}
=
\frac{1}{2}\sum_{m=0}^n \binom{n}{m}
\parbox{200pt}{
\begin{fmfgraph*}(160,60)
\fmfpen{1pt}
\fmfincoming{i1,i2,i3,i4,i5}
\fmfoutgoing{o1,o2,o3,o4,o5}
\fmfv{d.sh=circle,d.size=30pt,d.filled=empty}{v1}
\fmfv{d.sh=circle,d.size=30pt,d.filled=empty}{v2}
\fmf{phantom}{i2,h1,hh1,v1,hh2,h2,i3}
\fmf{phantom}{i4,h3,hh3,v1}
\fmf{phantom}{v2,hh4,h4,o2}
\fmf{phantom}{o3,h5,hh5,v2,hh6,h6,o4}
\fmfdotn{h}{6}
\fmf{dashes}{i1,v1,i5}
\fmf{dashes}{o1,v2,o5}
\fmf{plain,tension=2.5}{v1,u1}
\fmf{plain,tension=2.5}{u1,v2}
\fmfv{d.sh=circle,d.size=12pt,d.filled=shaded}{u1}
\fmflabel{$1$}{i5}
\fmflabel{$m$}{i1}
\fmflabel{$n$}{o5}
\fmflabel{$m+1$}{o1}
\end{fmfgraph*}
}
\end{gather*}

\end{fmffile}

\section*{Acknowledgements}
Thanks to J\"org Lemm, Gernot M\"unster, Martin Rehwald, and 
Manfred Salmhofer for pedagogical advice.

\typeout{}
\typeout{Please type "mp graph" and rerun LaTeX to produce 
the Feynman graph.}
\typeout{}


\begin{thebibliography}{X}
\bibitem{Zinn-Justin:1990} J.~Zinn-Justin, Quantum Field Theory 
and Critical Phenomena, 
Oxford Science Publications 1990
\bibitem{Coleman:1973} S.~Coleman, Laws of Hadronic Matter, 
Erice 1973
\bibitem{Brydges:1992} D.~Brydges, Functional Integrals and 
their Applications, Lausanne 1992
\bibitem{Brydges/Yau:1990} D.~Brydges, H.~Yau, 
Grad $\phi$ Perturbations of Massless Gaussian Fields,
Commun.~Math.~Phys.~129, 351-392 (1990)
\end{thebibliography}
\end{document}